\documentstyle[aps,eqsecnum,graphicx,preprint]{revtex}

\newcommand{\bm}[1]{{\rm \bf #1}}
\newcommand{\bp}{{\rm \bf p}}
\newcommand{\bk}{{\rm \bf k}}

\begin{document}
\draft
\title{The Conductivity Tensor for the Hubbard Model}
\author{F. Mancini$^{a}$ and D. Villani$^{b}$\footnote{Corresponding author.
E-mail address: dvillani@physics.rutgers.edu}}
\address{$^{a}$ Dipartimento di Scienze Fisiche ``E.R. Caianiello" e
Unit\`a I.N.F.M. di Salerno \\ Universit\`a di Salerno, 84081 Baronissi
(SA), Italy}
\address{$^{b}$ Serin Physics Laboratory, Rutgers University,
Piscataway, New Jersey 08855-0849} \maketitle
\begin{abstract}
A new theoretical analysis of the current-charge and charge-charge
propagators is presented for the Hubbard model, using the static
approximation for the Composite Operator Method. This approximation
manifestly preserves a sum rule which governs the single-site dynamics.
We compare our results with those obtained by numerical analysis.

\noindent{\bf Keywords:} Drude weight, Ward-Takahashi identities,
Hubbard model. \pacs{72.10.-d, 71.10.Fd, 74.72.-h, 71.10.-w}
\end{abstract}

{\bf 1.} The study of optical spectra is an important probe of the
dynamics of strongly correlated electronic systems and gives
information about the normal and superconducting states of these
materials \cite{1,2}. Also, the optical conductivity is of fundamental
theoretical interest because the spectral weight at low frequencies
might be a natural order parameter for the Mott transition \cite{3}.

The optical spectrum of undoped cuprate compounds, such as $La_2CuO_4$,
$YBa_{2}Cu_{3}O_{6}$ and $Nd_{2}CuO_{4}$, is characterized by the
$O2p-Cu3d_{x^{2}-y^{2}}$ charge-transfer excitation with a threshold
energy in the $1.5-2.0eV$ range \cite{1,4}. By doping charge carriers,
a drastic change in the optical spectrum occurs. The charge-transfer
excitation gap is rapidly suppressed and instead an edge shows up
around $1.0eV$ in the reflectivity spectrum, indicating that the
spectrum is dominated by low-energy excitations in the energy range
lower than $1.0eV$ \cite{5}. This kind of changes with carrier doping
are universal for all known hole-doped and electron-doped cuprate
superconductors. Besides, the appearance of optical anomalies is not
unique to the doped copper oxides, but is almost universally observed
in the spectra of the systems with strong electron correlation
\cite{1,2,8}.

In the simplest form, the Hubbard model is believed to describe many
properties of strongly correlated fermion systems. The applicability of
the model to the superconducting copper-oxides is related to the fact
that lower and upper Hubbard subbands can mimic the $O2p$ and
$Cu3d_{x^{2}-y^{2}}$ bands of the cuprates and thus the charge-transfer
gap can be described by an effective Hubbard on-site coupling constant.

Theoretical studies of the optical conductivity for the Hubbard model
started long time ago \cite{10}: it was investigated by the moment
method, by the equation of motion method for the Green's functions in
the Hubbard I approximation and in the Hubbard III approximation. For
the two-dimensional Hubbard model the strong coupling limit has been
considered mostly by numerical methods based on exact diagonalization
for small clusters \cite{11,12,13,14,15,16,17}. A detailed discussion
of optical and photoemission sum rules for one- and two- dimensional
Hubbard model has been given recently by comparing a perturbation
theory in the kinetic term with numerical calculations \cite{18}.

An analytical analysis of the frequency dependent conductivity has been
given by perturbation method in weak-coupling limit \cite{19,20} and by
applying the memory function technique in terms of the Hubbard
operators \cite{10}. Recently the optical conductivity has been
investigated within the dynamical mean-field approach [6]. However, for
the realistic planar Hubbard model non-local corrections to the
transport vertices and the self-energy are important [6]. Beyond this,
vertex corrections are crucial to satisfy the relevant Ward-Takahashi
identities [17].

In this Letter, we present a new theoretical analysis of the
conductivity tensor by use of the Composite Operator Method [18] for
the two-dimensional Hubbard model. By exploiting sum rules after the
Ward-Takahashi identities and the Pauli principle, the calculation of
the current-current propagator is reduced to the charge-charge one in
the framework of a fully self-consistent scheme. Further, it will be
shown that the charge-charge propagator preserves an important sum rule
which governs the single-site dynamics avoiding double occupancy
configurations unless to form local singlets.

The Letter is organized as follows. First, the model is set up and the
formula for the conductivity tensor in the linear response theory is
summarized. The Ward-Takahashi identities are given in the context of
the two-dimensional Hubbard model and the hierarchy degrading the
current-current propagator to the current-charge and charge-charge ones
is built up. The calculation of the current-charge and charge-charge
propagators is realized within the static approximation for the
Composite Operator Method. Then, the results for the optical
conductivity and the Drude weight are presented and compared with those
obtained by numerical analysis of small clusters. Some concluding
remarks are presented at the end.

{\bf 2.} In the presence of an electromagnetic field described by the
vector potential $\vec A(\bm{r},t)$, the Hubbard model is as follows
\begin{equation}
H=\sum_{\langle ij\rangle} \left[ t_{ij} (A) - \mu \delta_{ij}\right]
c^\dagger (i) c(j) +U \sum_i n_\uparrow (i) n_{\downarrow}(i)
\end{equation}         
where we use a standard notation. The hopping matrix elements contain
the Peierls factors which guarantee the gauge invariance.

In the framework of the linear response theory, the electric
conductivity tensor is given as follows
\begin{equation}
\sigma_{ab}(\bm{k},\omega) = e^{2 }{1\over i(\omega+i \eta)}\langle
K_{a}\rangle \delta_{ab}-{1\over i(\omega+i \eta)}g_{ab}(\bm{k},
\omega)
\end{equation}
where $g_{ab}(\bm{k}, \omega)$ is the retarded current-current
propagator, with $K_a(i)$ being the kinetic energy density operator in
the $a$ direction.

In particular, the frequency-dependent uniform, i.e. $\bm{k}=\bm{0}$,
electric conductivity $\sigma_{xx}(\omega)$ is given by
\begin{equation}\label{2.14}
\sigma_{1}(\omega) ={\rm Re} \sigma_{xx}(\omega) =D\delta(\omega) +
\sigma _{inc}(\omega)
\end{equation}
where $D$ is the Drude weight
\begin{equation}
D=e^{2}\pi \left[-\langle K_{x}\rangle +{1\over e^{2}} {\rm Re}
g_{xx}(\bm{0},\omega\to 0)\right]
\end{equation}       
which measures the ratio of the density of the mobile charge carriers
to their mass [3], and $\sigma _{inc}$ is the incoherent contribution
\begin{equation}
\sigma_{inc}(\omega) =-{1\over \omega}{\rm Im} g_{xx}(\bm{0},\omega)
\end{equation}

By defining the charge $\rho(i)$ and current $\bm{j}(i)$ densities as
\begin{eqnarray}
\rho(i) &=& ec^\dagger(i) c(i) \\ 
\bm{j}(i) &=& {tea^{2}\over i} c^\dagger(i) \left[\overrightarrow\nabla
-
\overleftarrow\nabla \right] c(i)
\end{eqnarray}
it is immediate to obtain the conservation law
\begin{equation}\label{3.5}
\nabla \cdot \bm{j} (i) +{\partial\over \partial t}\rho (i) =0
\end{equation}
The symmetry content of the algebraic equation (\ref{3.5}) manifests at
level of observation as relations among matrix elements once a choice
of the physical space of states has been made. Indeed, by defining the
causal charge and current propagators as $\chi_{ab}(i,j) = \langle
T\left[ g_{a }(i) g_{b}(j) \right]\rangle$ where
$g_{a}(i)=\left(\rho(i),\,j_{x}(i),\,j_{y}(i)\right)$ for
$a=0,\,x,\,y$, we can derive a series of Ward-Takahashi identities,
which in momentum space read as
\begin{eqnarray}
ia\omega\chi_{00}(\bm{k},\omega) & =& \left[1-e^{-ik_{x}a}\right]
\chi_{xo}(\bm{k},\omega)+
\left[1-e^{-ik_{y}a}\right]\chi_{y0}(\bm{k},\omega)\\
ia\omega\chi_{x0}(\bm{k},\omega) & =& -2te^2a^2\langle c^\dagger(i)
c^\alpha(i) \rangle \left[ 1-e^{ik_{x}a}\right] \nonumber\\ && -
\left[1-e^{ik_{x}a}\right] \chi_{xx} (\bm{k},\omega) -
\left[1-e^{ik_{y}a}\right] \chi_{xy}
(\bm{k},\omega) \\     
 ia\omega\chi_{y0}(\bm{k},\omega) & = & -2te^2a^2\langle c^\dagger(i)
c^\alpha(i) \rangle \left[ 1-e^{ik_{y}a}\right] \nonumber\\ && -
\left[1-e^{ik_{x}a}\right] \chi_{yx} (\bm{k},\omega) -
\left[1-e^{ik_{y}a}\right] \chi_{yy}
(\bm{k},\omega) \\     
a^{2}\omega^{2} \chi_{00} (\bm{k},\omega) &=& 8ta^{2} e^{2}
\left[1-\alpha(\bm{k}) \right] \langle c^\dagger (i) c^\alpha(i)
\rangle \nonumber\\ && +2\chi_{xx} (\bm{k},\omega)\left[ 1-\cos
(k_{x}a)\right] + \chi_{xy}(\bm{k},\omega) \left[1-e^{-ik_{x}a}\right]
\left[1-e^{ik_{y}a}\right] \label{3.11}\\ 
 && + \chi_{yx}(\bm{k},\omega) \left[1-e^{ik_{x}a}\right]
 \left[1-e^{-ik_{y}a}\right] +2 \chi_{yy} (\bm{k},\omega)\left[1-\cos
(k_{y}a)\right]
\end{eqnarray}
By these identities we see that the current-current propagators
$\chi_{ab}(\bm{k},\omega)$ can be expressed in terms of the
current-charge and the charge-charge ones.

In order to calculate the current-charge and the charge-charge
propagators, we use the static approximation [19] for the composite
field
\begin{equation}
\psi(i) = \left(\begin{array}{c} \xi(i)\\ \eta(i)
\end{array}\right)
\end{equation}  
By defining
\begin{equation}
N(k) = F.T. \langle T[n(i) n(j) ]\rangle \qquad S(k) = F.T. \langle
T[\psi(i) \psi^\dagger (j) ]\rangle
\end{equation} 
\begin{equation}
I=F.T. \langle \left\{ \psi(i) ,\psi^\dagger (j) \right\}_{E.T.}\rangle
= \left( \begin{array}{cc} 1-{n\over 2} & 0\\ 0 & {n\over 2}
\end{array} \right)
\end{equation} 
we have
\begin{eqnarray}
N(k) = -in^{2}(2\pi)^{3}a^{-2} \delta^{(3)} (k) - {n(2-n)\over n-2D}
&&\left\{ I^{-1}_{11}\left[ Q_{1111}(k) + Q_{1112}(k)+Q_{1211}(k) +
Q_{1212}(k)\right] \right.\nonumber\\ && + \left. I^{-1}_{22}\left[
Q_{1212}(k) + Q_{1222}(k)+Q_{2212}(k) + Q_{2222}(k)\right]\right\}
\end{eqnarray}  
with
\begin{equation}
Q_{\alpha\beta\gamma\delta}(\bk,\omega)={i\Omega\over (2\pi)^{3}}\int
d^2 pd\Omega S_{\alpha\beta} (\bk+\bp,\omega+\Omega) S_{\gamma\delta}
(\bp,\Omega)
\end{equation}
where D is the double occupancy
\begin{equation}
D={1\over 2} \langle \eta^\dagger (i) \eta (i) \rangle
\end{equation} 
$k$ is the four-dimensional vector $k=(\bk, \omega)$; the symbol F.T.
denotes the Fourier transform.

Then, it is direct to see that
\begin{equation}
\chi_{00}(k) =e^{2}N(k)
\end{equation} 
and
\begin{eqnarray}\label{4.20}
\chi_{a0}(k) &=& (-ite^{2}a) (e^{ik_{a}a}-1) N (k) \left\{
I^{-1}_{11}\left[ \langle \xi^\alpha \xi^\dagger \rangle + \langle
\xi^\alpha \eta^\dagger\rangle \right] - I^{-1}_{22} \left[ \langle
\eta^\alpha \xi^\dagger \rangle + \langle \eta^\alpha
\eta^\dagger\rangle \right] \right\} \nonumber\\ &&-2 I^{-1}_{11}
(-ite^{2}a) \left[ Q^a_{1111}(k) + Q^a_{1112}(k) + Q^a_{1211}(k) +
Q^a_{1212}(k)\right]\nonumber\\ && -2 I^{-1}_{22} (-ite^{2}a) \left[
Q^a_{1212}(k) + Q^a_{1222}(k) + Q^a_{2212}(k) + Q^a_{2222}(k)\right]
\end{eqnarray} 
where we have defined
\begin{equation}\label{4.21}
Q^a_{\alpha\beta\gamma\delta}(\bk,\omega)={i\Omega\over (2\pi)^{3}}\int
d^2 pd\Omega \left[e^{i(p_{a}+k_{a})a} -e^{-ip_{a}a}\right]
S_{\alpha\beta} (\bk,\Omega) S_{\gamma\delta} (\bp+\bk,\Omega+\omega)
\end{equation}  
We have in the coordinate space
\begin{eqnarray}\label{4.22}
N(i,j) &=& n^2 - {n(2-n)\over n-2D} \left\{
I^{-1}_{11}\left[Q^a_{1111}(i,j) + Q^a_{1112}(i,j) + Q^a_{1211}(i,j) +
Q^a_{1212}(i,j)\right]\right.\nonumber\\ && \left.+ I^{-1}_{22} \left[
Q^a_{1212} (i,j)+ Q^a_{1222}(i,j) + Q^a_{2212}(i,j) +
Q^a_{2222}(i,j)\right] \right\}
\end{eqnarray} 
In particular, at equal sites (\ref{4.22}) reduces to
\begin{equation}
N(i,i) = \langle n(i) n(i)\rangle = n+2D
\end{equation} 
We thus see that the charge propagator satisfies the sum rule
\begin{equation}\label{4.24}
{ia^2\over (2\pi)^{3}} \int d^{2}kd \omega \chi_{00}(k) = e^{2} (n+2D)
\end{equation} 
as it should be. The physical content of (\ref{4.24}) is that of the
Pauli principle because it governs the single-site dynamics of two
fermions.

By means of the Ward-Takahashi identity (\ref{3.11}) we obtain
\begin{equation}\label{5.1}
\chi_{xx}(k_{x}, k_{y} = 0, \omega) = {a^{2} \omega^{2} \chi_{00}(
k_{x}, k_{y}=0, \omega) \over 2\left[1-\cos(k_{x}a)\right]} +{1\over 2}
a^{2}e^{2}K
\end{equation} 
where $K=-4t\langle c^\dagger (i) c^\alpha(i)\rangle$ is the kinetic
energy per site. From the analytical structure of equation (\ref{5.1}),
we note that $\chi_{xx}(k_{x},k_{y}=0,0)=a^{2}e^{2}K/2$ does not depend
on $k_{x}$ and is proportional to the kinetic energy. It is remarkable
that such a peculiar behaviour has been previously found by use of
Quantum Monte Carlo techniques [8]. The correspondence with the
notation of Ref. 8 is given by $g_{xx}=-e^{2} \Lambda _{xx}$.

By taking the limit $k_{x}\to 0$ in (\ref{5.1}) and by considering the
retarded function we have
\begin{equation}
g_{xx}(0,\omega)={1\over 2} a^2 e^{2} K + \lim_{k_{x}\to
0}{a^{2}\omega^{2}\chi^{R}_{00}(k_{x},\omega)\over
2\left[1-\cos(k_{x}a)\right]}
\end{equation} 
Recalling (\ref{2.14}) we have for the optical conductivity
\begin{equation}
\sigma_{1}(\omega) =D\delta(\omega) + \sigma _{inc}(\omega)
\end{equation}  
where the Drude weight D is given by
\begin{equation}
D=\pi{\rm Re} \lim_{\omega\to 0}\lim_{k_{x}\to
0}{a^{2}\omega^{2}\chi^{R}_{00}(k_{x},\omega)\over
2\left[1-\cos(k_{x}a)\right]}
\end{equation} 
and the incoherent part $\sigma _{inc}(\omega)$ is as follows
\begin{equation}
\sigma _{inc}(\omega)=-{1\over \omega} {\rm Im} \lim_{k_{x}\to
0}{a^{2}\omega^{2}\chi^{R}_{00}(k_{x},\omega)\over
2\left[1-\cos(k_{x}a)\right]}
\end{equation} 
As emphasized in the beginning, the whole problem of the response to an
external electromagnetic field has been reduced to the evaluation of
the charge-charge propagator. In the static approximation of the
Composite Operator Method [18-20] the charge-charge propagator, as the
spin-spin one, can be connected to convolutions of single-particle
propagators [19]. It is important to note that the convolutions in
(\ref{4.20}) and (\ref{4.21}) involve higher order single-particle
Green's functions. Then, such a scheme does not contradict the
conclusion reached by exact diagonalization analysis [11] that any
attempt to compute the conductivity from a convolution of
single-particle propagators is destined to fail in one dimension and is
questionable in planar systems. In our case, the occurrence of
convolutions is related to a linearized dynamics, but also involves
choice of occupation dependent electronic excitations as basic fields
[19]. In this way, the electron propagation is described as a
repetition of composite excitations which automatically take into
account scattering of electrons on spin and charge fluctuations due to
strong correlations. Also, the use of a higher order basic field gives
the advantage of implementing the Pauli principle and of a clear
theoretical understanding of the terms originating from intraband or
interband propagation. In fact, it turns out that only intraband
excitations contribute to the Drude weight, whereas interband ones
build up the incoherent part.

In the static approximation, by means of the previous results, the
retarded charge-charge propagator is given by
\begin{eqnarray}
\chi^{R}_{00}(\bk,\omega) &=& -in^{2}e^{2}(2\pi)^{3}a^{-2}\delta^{(3)}
(k) - {n(2-n)e^{2}\over n-2D} \left\{I^{-1}_{11} \left[ Q^R_{1111}(k) +
Q^R_{1112}(k) \right.\right.\nonumber\\ && \left.\left.+ Q^R_{1211}(k)
+ Q^R_{1212}(k)\right] + I^{-1}_{22} \left[ Q^R_{1212}(k) +
Q^R_{1222}(k) + Q^R_{2212}(k) + Q^R_{2222}(k)\right] \right\}
\end{eqnarray}
where
\begin{equation}
Q^R_{\alpha\beta\gamma\delta}(\bk,\omega)=\sum^{2}_{i,j=1} {\Omega\over
(2\pi)^{2}}\int d^2 p {f[E_{j}(\bk+\bp)]-f[E_{i}(\bp)]\over \omega +
E_{i}(\bp) - E_{j}(\bk+\bp)+i\eta} \sigma^{(i)}_{\alpha\beta}(\bp)
\sigma^{(j)}_{\gamma\delta}(\bk+\bp)
\end{equation}  
$f(\omega)$ is the Fermi distribution function.
$\sigma^{(i)}_{\alpha\beta}(\bp)$ are the spectral functions of the
single-particle propagator. These functions have been previously
calculated [18, 19].

Let us define
\begin{eqnarray}
A_{ij}(\bk,\bp) &=& I^{-1}_{11}\left[ \sigma^{(i)}_{11}(\bp)
\sigma^{(j)}_{11}(\bk+\bp) + \sigma^{(i)}_{11}(\bp)
\sigma^{(j)}_{12}(\bk+\bp) + \sigma^{(i)}_{12}(\bp)
\sigma^{(j)}_{11}(\bk+\bp) + \sigma^{(i)}_{12}(\bp)
\sigma^{(j)}_{12}(\bk+\bp)\right]\nonumber\\ &&+I^{-1}_{22}\left[
\sigma^{(i)}_{12}(\bp) \sigma^{(j)}_{12}(\bk+\bp) +
\sigma^{(i)}_{12}(\bp) \sigma^{(j)}_{22}(\bk+\bp) +
\sigma^{(i)}_{22}(\bp) \sigma^{(j)}_{12}(\bk+\bp) +
\sigma^{(i)}_{22}(\bp) \sigma^{(j)}_{22}(\bk+\bp)\right]
\end{eqnarray} 
and
\begin{equation}
X_{ij}(\bk,\omega)={\Omega\over (2\pi)^{2}}\int d^2 p
{f[E_{j}(\bk+\bp)]-f[E_{i}(\bp)]\over \omega + E_{i}(\bp) -
E_{j}(\bk+\bp)+i\eta} A_{ij}(\bk,\bp)
\end{equation} 
Then, $\chi^{R}_{00}(\bk,\omega) $ can be rewritten as
\begin{equation}
\chi^{R}_{00}(\bk,\omega) = -in^{2}e^{2}(2\pi)^{3}a^{-2}\delta^{(3)}
(k) - {n(2-n)e^{2}\over n-2D}\sum^{2}_{i,j=1} X_{ij}(\bk,\omega)
\end{equation}
and
\begin{equation}
D=- {n(2-n)\pi e^{2}\over n-2D}\sum^{2}_{i,j=1} \lim_{\omega\to
0}\lim_{k_{x}\to 0}{a^{2}\omega^{2}{\rm Re}X_{ij} (k_{x},\omega)\over
2\left[1-\cos(k_{x}a)\right]}
\end{equation} 
It is direct to see that the interband terms $X_{12}(\bk,\omega)$ and
$X_{21}(\bk,\omega)$ do not contribute. Also, by means of
straightforward calculations the contribution of the intraband terms
gives
\begin{equation}
 \lim_{\omega\to 0}\lim_{k_{x}\to 0}{a^{2}\omega^{2}{\rm Re}X_{ij}
(k_{x},\omega)\over 2\left[1-\cos(k_{x}a)\right]} = -{a^{2}t^{2}\over
4k_{B}T}{\Omega\over (2\pi)^{2}}\int d^{2}p\sin^{2} (ap_{x})X_{i}(\bp)
\end{equation} 
and the Drude weight takes the expression
\begin{equation}
D={n(2-n)\pi e^{2}a^{2}t^{2}\over 4k_{B}T (n-2D)}\sum^{2}_{i=1}
{\Omega\over (2\pi)^{2}}\int d^{2}p\sin^{2}(ap_{x})X_{i}(\bp)
\end{equation} 
where we defined
\begin{equation}
X_{i}(\bp) = \epsilon^{2}_{i}(\bp) [1-T^{2}_{i}(\bp)] A^{(0)}_{ii}(\bp)
\end{equation} 
with
\begin{equation}
T_{i}(\bp) = \tanh \left( {E_{i}(\bp) - \mu \over 2k_{B}T}\right)
\end{equation} 

In the same way, it is straightforward to show that the incoherent part
is given by
\begin{equation}
\sigma _{inc}(\omega)={1\over \omega} {n(2-n) e^{2}\over
n-2D}\sum^{2}_{i,j=1} {\rm Im} \lim_{k_{x}\to 0}
{a^{2}\omega^{2}X_{ij}(k_{x},\omega)\over 2\left[1-\cos(k_{x}a)\right]}
\end{equation} 
where the intraband terms do not contribute. In particular, it is found
that $\sigma _{inc}(\omega)=0$ for $\omega <\omega_{c}$. For any finite
value of $U$ there is a gap in the incoherent part of the optical
conductivity, given by
\begin{equation}\label{5.19}
\omega_{c}=2\sqrt{I_{11}I_{22}} U\qquad \hbox{[At half-filling
$\omega_{c}=U=2\mu$ ]}
\end{equation} 

In Fig. 1 we present results for the Drude weight and from numerical
analysis of small clusters [13].

For $U=15$ the Hubbard gap is clearly visible at half-filling because,
as discussed by Kohn [3], one expects that in an insulating phase $D$
will vanish. Also, there is no spectral weight in the interval $0\le
\omega\le 15$ [cfr. Eq. (\ref{5.19})]. Upon doping, spectral weight is
transferred from the region above the Hubbard gap to the Drude peak at
zero frequency. This is up to a critical doping where there is a
deviation downwards which may be taken as a signature of a drastic
change in the nature of the carriers. Indeed, after this doping the
system starts to resemble a gas of non-interacting electrons and the
Drude weight follows the kinetic energy [12]. It is worth noting that
the value of the Drude weight and of its slope in proximity of
half-filling should be carefully interpreted within the scheme used. In
fact, many numerical and theoretical indications [2] signal the onset
of additional states in the gap due to doping, extending upwards from
the lower Hubbard band, which absorb a large spectral weight. Then, the
absence of a mid-gap band accompanied by a conserved total spectral
weight, as in the approximation used [20], will very much affect the
spectral weight transfer between the interband charge excitations and
the intraband ones. Further, this occurrence can be responsible of a
Drude weight larger then the non-interacting value near half-filling
for some particular choice of the on-site Coulomb repulsion $U$.

The kinetic energy is shown in Fig. 2. The results have been obtained
through the scheme given in Refs. 21 and 22 which makes extensive use
of thermodynamic identities for the Helmholtz free energy per site.

These results to a large extent contradict data available both from
numerical and experimental analysis [1, 2]. The observed ratio
$\omega_{c}/U$, connected to the energy scale associated with interband
charge excitations, is greater compared to what is expected from
Lanczos's techniques [12]. $\omega_{c}/U$ is depicted as function of
the filling $n$ in Fig. 3.

However, the small system sizes and the lack of theoretical analysis of
finite size corrections make interpretation of the results uncertain. A
more troublesome discrepancy is related to the absence of a clear
feature centered at low frequency beyond the Drude peak, quickly
developing by varying the dopant concentration, as already discussed in
relation with the Drude weight results. The presence of a mid-gap band
is a genuine two-dimensional effect [11] that cannot be related to
interband charge excitations which lead to an absorption band centered
near $U$ , as seen also in the strong coupling counterpart of the
Hubbard model [2] (i.e., the $t-J$ model). This is the well known
mid-infrared band, which has been observed in several cuprate
superconductors and strongly correlated electronic systems by use of
analytical and numerical techniques [1, 2]. We do not at present have a
resolution of this discrepancy, but a possible one is as follows.

The energy scale on which the center of gravity of the mid-infrared
band is centered is $4t^{2}/U$, that is the value of the Heisenberg
exchange $J$ when derived from the one-band Hubbard model. Then,
optical mid-gap effects can be observed only if the scheme presents in
addition to the lower Hubbard band a quasi-particle peak separated by a
gap or pseudogap of order $J$. On the contrary, we cannot reproduce the
latter result since for our estimations of the single-particle
propagator we have used a two-pole expansion for the one-electron
spectral density. In this respect, an enlarged composite basis with a
third asymptotic excitation can resolve this structure.

In conclusion, we have shown that the operatorial equation (\ref{3.5})
expressing the charge conservation law has a direct corollary in terms
of the existence of Ward-Takahashi identities, which hierarchically
degrade the evaluation of the current-current propagator to the
current-charge and charge-charge ones. Such an occurrence takes place
also in the non interacting system (i.e., $U=0$). However, it is worth
pointing out the different role played by the Ward-Takahashi identities
in non interacting systems compared to that in the framework of an
approximate theory for interacting ones. In the first case the
Ward-Takahashi identities are automatically satisfied and it is a
matter of convenience to compute the conductivity from the
current-charge or charge-charge propagators. In a theoretical scheme
for an interacting system there is no way to avoid some approximation.
Then, in addition to a clear convenience in computing a hierarchically
degraded propagator, there is the possibility to build up a physical
space of states where the general symmetry principle of charge
conservation law can be enforced to model the dynamics. The scheme
turns out to be even more coherent by preserving the fermionic
character of the propagating charges or, in other words, by
implementing the Pauli principle. That is, by recovering such general
symmetries the propagation of the charge and their intrinsic dynamics
is constrained in a suitable Hilbert space where the charges propagate
without degrading in number and avoiding double occupancy
configurations unless to form local singlets.

The authors would like to express their thanks to Adolfo Avella for
valuable discussions and Sarma Kancharla for a careful reading of the
manuscript. One of us (D.V.) thanks Gabriel Kotliar for kind
hospitality during his stay at the Rutgers University where a part of
this work has been done.

\begin{figure}
\begin{center}
\includegraphics*[width=7.5cm]{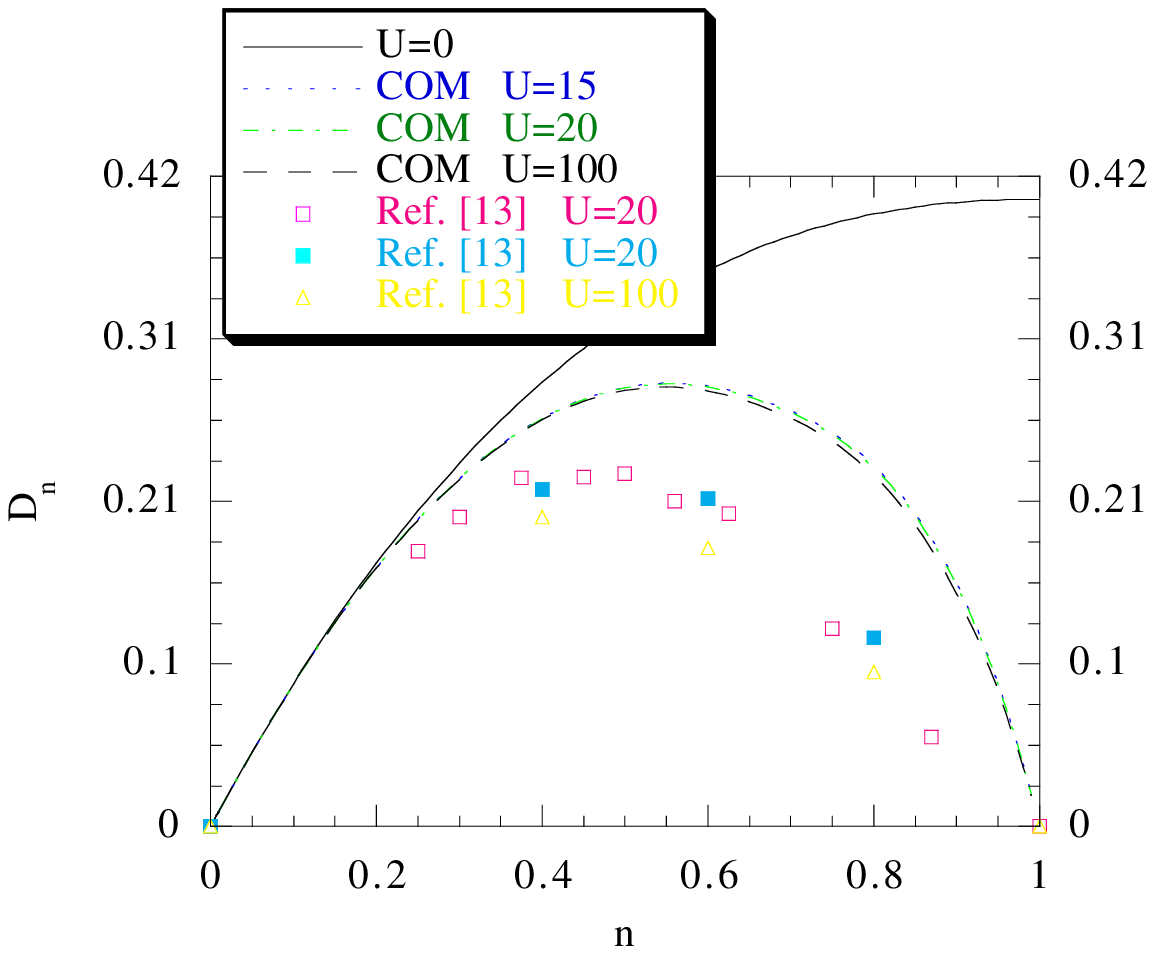}
\end{center}
\caption{ The normalized Drude weight $D_{n}=D/(2\pi e ^{2})$ as a
function of the filling $n$ for various couplings $U$. Full squares are
zero temperature data using exact diagonalization techniques on
$4\times 4$ sites. Open squares and triangles indicate results for a
$\sqrt{10}\times \sqrt{10}$ site cluster.}
\end{figure}

\begin{figure}
\begin{center}
\includegraphics*[width=7.5cm]{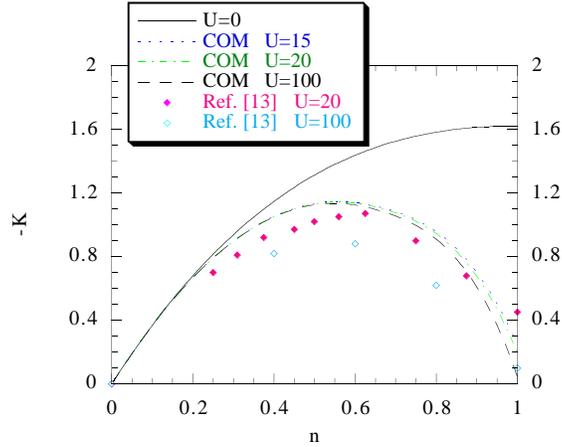}
\end{center}
\caption{Kinetic energy per site $-K$ as a function of the filling $n$
for various couplings $U$. Full rhombi are zero temperature data using
exact diagonalization techniques on $4\times 4$ sites. Open rhombi
indicate results for a $\sqrt{10}\times \sqrt{10}$ site cluster.}
\end{figure}

\begin{figure}
\begin{center}
\includegraphics*[width=7.5cm]{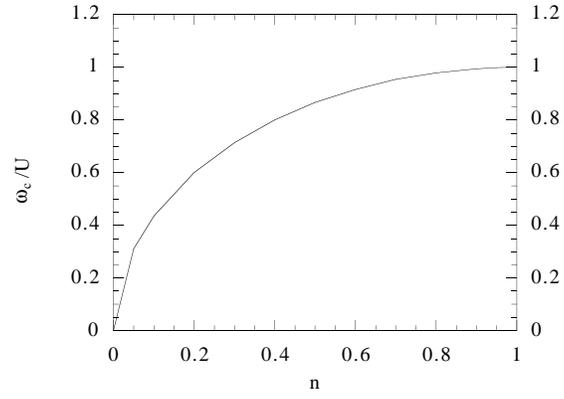}
\end{center}
\caption{$\omega_{c}/U$ as function of the filling $n$.}
\end{figure}

\end{document}